\documentclass[twocolumn,showpacs,preprintnumbers,amsmath,amssymb,floatfix]{revtex4}

\usepackage{graphicx}
\usepackage{epsfig} 
\usepackage{dcolumn}
\usepackage{bm}
\usepackage{epstopdf}
\usepackage{multirow}
\usepackage{amsmath}
\usepackage{booktabs}
\usepackage{color}

\begin{document}
	
\preprint{APS/123-QED}
	
\title{Genetic fitting techniques for precision ultracold spectroscopy}
	
\author{I. C. Stevenson,$^{1}$, J. P\'{e}rez-R\'{i}os$^{2, \ 3}$}
\affiliation{%
	$^1$School of Electrical and Computer Engineering Purdue University, $^2$Department of Physics and Astronomy Purdue University, $^3$School of Natural Sciences and Technology, Universidad del Turabo
}
	
\date{\today}
	
\begin{abstract}
We present development of a genetic algorithm for fitting potential energy curves of diatomic molecules to experimental data. Our approach does not involve any functional form for fitting, which makes it a general fitting procedure. In particular, it takes in a `guess' potential, perhaps from an \textit{ab initio} calculation, along with experimental measurements of vibrational binding energies, rotational constants, and their experimental uncertainties. The fitting procedure is able to modify the guess potential until it converges to better than 1\% uncertainty, as measured by $\bar{\chi}^2$.  We present the details of this technique along with a comparison of potentials calculated by our genetic algorithm and the state of the art fitting techniques based on inverted perturbation approach for the $X \ ^1\Sigma^+$ and $C \ ^1\Sigma^+$ potentials of lithium-rubidium.
\end{abstract}

\maketitle

\section{Introduction}
The cornerstone of modern chemistry is the study of molecular interactions and how these lead to all chemical reactions governing the evolution of the universe. However, the available experimental tools, such as spectroscopy and scattering observables, do not generate direct measurements of these interactions, and therefore mapping techniques that translate collisional and spectroscopic data into the underlying interactions are necessary for the chemistry and physics community. First studied in the 1930's, molecular spectral line broadening revealed the long range interactions between the molecules~\cite{Kuhn1934,Kuhn1937,Kuhn1937bis}. Surprisingly, this approach is still useful for spectroscopy of Rydberg excitations in dense background gases at ultracold temperatures~\cite{JPR2016,Michael2016,Tara2016}. From there, better techniques have resulted in more accurate data. Modern spectroscopic techniques, such as Fourier-transform spectroscopy~\cite{becker1972fourier}, photoassociation of ultracold atoms~\cite{thorsheim1987laser}, and measurement of Feshbach resonances~\cite{Chin2010}, have led to the most accurate atom-atom interaction potentials to date. In the same vein, thanks to the development of molecular beam technology, similar techniques transform scattering information of atom-molecule and molecule-molecule collisions into realistic atom-molecule and molecule-molecule potential energy surfaces,~\cite{Aquilanti1998,Aquilanti1999,Aquilanti2002,Bernstein1973,Bernstein1967,Cappelletti2005,Greene1972} although the dynamics require special care~\cite{JPR2012}.

Several fitting or inversion techniques to transform vibrational, rotational and hyperfine spectroscopic data into molecular interactions have been developed; semi-classical techniques lead to the celebrated LeRoy-Bernstein formula~\cite{leroy1970dissociation} and the RKR method (Rydberg, Klein and Rees)~\cite{Rees1947}. For more accurate results, the inverted perturbation approach (IPA)~\cite{Hinze1975,Vidal1977} modifies the RKR result by solving the Schr\"odinger equation and adjusting the potential to better reproduce the experimental data.  Many of the older techniques, share something in common: the potential can be found if a functional form is assumed. A functional form limits the search space of the optimization routine and prevents it converging to a true solution if nature escapes description by a simple analytic function. However, more modern techniques such as some variants of the IPA method~\cite{ivanova2011x} and machine learning techniques, and in particular genetic algorithms (GA), use non-functional based methods and the machine learning techniques present an alternative to standard fitting procedures with higher accuracy and a more general mathematically fundamental basis.

\begin{figure*} [t]
	\includegraphics[width=\textwidth]{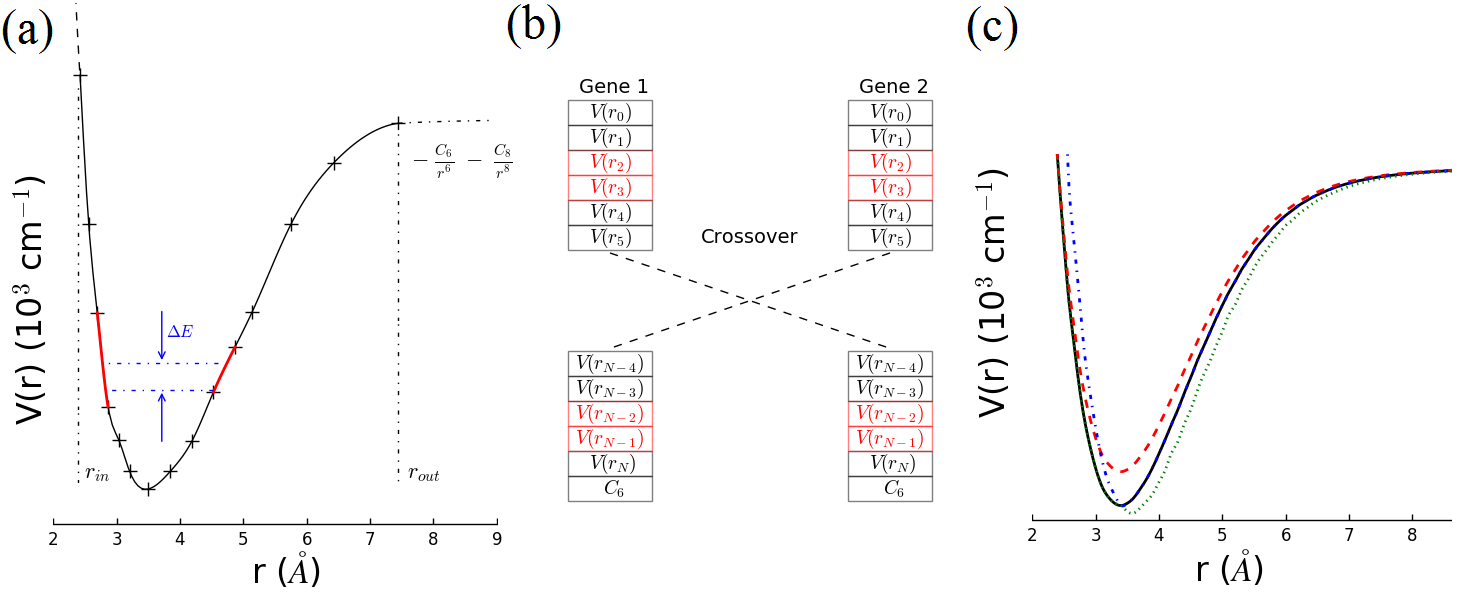}
	\caption{\textbf{(a)} Potential representation.  We represent the potential as three disjoint pieces: a short range piece for $r<r_{in}$ (dashed line), a point-wise potential connected by cubic splines for $r_{in}<r<r_{out}$ (solid line), and a long range piece for $r_{out}<r$ given by Eq.~(\ref{Eq:lrp}) (dot-dashed line).  The blue dot-dashed lines in the potential represent two vibrational levels with an energy difference $\Delta E$.  The highlighted red potential sections are nearby points in the point-wise representation that set the slope of the potential and determine $\Delta E$.  \textbf{(b)} Crossover operator and encoding of potentials onto genes.  An array stores the energies of the point-wise potential on a common $r$-space grid, with C$_6$ as the last entry.  Crossover happens at a random point in the array, shown here for the 6$^{\rm th}$ position, with the potentials swapping information on the inner and outer walls.  The entries in red correspond to the highlighted red region in \textbf{(a)}.  Because the slope of the potential is local to the nearest points, the crossover operation usually preserves the slope making $\Delta E$ a useful observable in our fitting procedure.  \textbf{(c)} Effect of specialized mutation operators on the potential.  These help correct systematic biases in the starting potential.  The black solid line is the starting potential; the red dashed line is the result of changing the well depth; the blue dot-dashed line is the result of changing the inner-wall location; and the green dotted line is the result of changing the outer-wall location.}
	\label{fig:geneEncoding}
\end{figure*}

Development of genetic algorithms~\cite{holland1992} runs parallel to development of inversion techniques for spectroscopy.  These non-linear optimization programs solve complex problems by mimicking biological reproduction and evolution. The algorithm encapsulates all the fitting information into a gene, analogous to a biological system, where the genes contain all the information for a living organism. As with real genetics, a genetic algorithm uses crossover, mutation and selection to descend from a starting initial population of genes and reaches a final answer, under some fitness criteria, \textit{i.e.}, it finds the best adapted gene. Perhaps surprisingly, genetic algorithms are a very powerful technique for solving hard numeric problems, and variants on the classic genetic algorithm are numerous. 

In the present paper, we present a novel fitting procedure based on genetic techniques to obtain molecular potentials from spectroscopic data. The method is function-free, {\it i.e}, it does not require any particular set of functions for the fitting, which translates to a more general approach leading to accurate potentials, and only requires an initial guess for the interaction. In particular, we fit two $^1\Sigma^+$ potentials in lithium-rubidium (LiRb) and compare our results with the state of the art fitting methods. Additionally, we speculate that this technique may find a home in inverting spectroscopic information of larger molecules, a burgeoning field for the ultracold community~\cite{morita2017cold} where a simple description of the interaction is impossible. The paper is structured as follows: in section 2 the fitting procedure is presented conceptually; in section 3 the accuracy of the technique is tested against the state-of-the-art {\it ab initio} results and the most advanced fitting techniques to date; in section 4 the technique is applied to a real world example; and finally in section 5 we present the conclusions and future work.

\section{Method} \label{sec:GA}

Choosing an appropriate potential representation and a method to encode it on a gene is the first step in implementing genetic based fitting for diatomic potentials, shown in Fig.~\ref{fig:geneEncoding} (a). We write the potential $V(r)$ as a function of the internuclear spacing $r$ separately in three regions: $r<r_{in}$; $r_{in}<r<r_{out}$; and $r_{out}<r$. For internuclear separation, $r<r_{in}$~\cite{footnote5} we extrapolate inwards from the last point in the point-wise potential with an exponential function as is customary for Morse type potentials. In the region $r_{in}<r<r_{out}$, a point-wise representation describes the potential, between which we interpolate with natural cubic splines. For $r>r_{out}$, a van der Waals interaction best describes the potential as
\begin{equation}
V(r) = V_{\infty} - \frac{C_6}{r^6} - \frac{C_8}{r^8},
\label{Eq:lrp}
\end{equation}
where $C_6$ is the van der Waals coefficient, $V_{\infty}$ is the dissociation energy, and $C_8$ is fitted for smooth connection to the point-wise potential. Fig.~\ref{fig:geneEncoding} (b) shows how we turn our potential representation into a gene for the genetic algorithm. We fix a grid in $r$-space for the point-wise potential to be shared among all genes. The genes hold the potential energy at each point in the $r$-space grid, $V(r_i)$, $i \ \in \ [0, \ N]$, where $N+1$ is the number of points in the point-wise potential, and the last entry in the gene list is $C_6$, which governs the long range part of the potential. 

With our potential stored as a gene, we turn our attention to the genetic operators: crossover, mutation and selection, which governs the descent to a final solution.  At the beginning of each generation, the fitness of each gene in the population is evaluated and given opportunities to reproduce based on how the gene scores relative to the population average.  For the first generation, our method uses the mutation operator(s) to create M variants on the guess potential, where M is the size of each generation.  From here reproduction is carried out with crossover and mutation to create a children population, whose fitness is also evaluated.  The children generation and parent generation are combined, and only the best M genes are kept, forming the parent genes for the next generation.  This sequence is iterated until the best gene reaches the desired accuracy and the program terminates.

A detailed discussion of each genetic operator is also warranted.  First, Fig.~\ref{fig:geneEncoding}~(b) graphically illustrates one point crossover, which allows an interchange of genetic information between genes and is the heart of a genetic algorithm.  A simple conceptual model for crossover is that it forms children with the `good' part of each parent gene with some probability. For our implementation of a GA we have several different types of mutation. Mutation helps mix in genes that are necessary, but which weren't included in the original population or which have been randomly eliminated by selection. We have a general mutation operator that modifies one point in the potential according to
\begin{equation}
V(r_i) = V_{\rm old}(r_i)(1+x),
\label{Eq:mutation}
\end{equation}
where $V_{\rm old}(r_i)$ is the old energy, $x$ is a random variable normally distributed with zero mean and variance $\sigma^2$.  Additionally we have several specialized mutation operators illustrated in Fig.~\ref{fig:geneEncoding} (c): changing the potential depth, moving the inner wall of the potential or moving the outer wall.  These operators exist because the potentials can have two types of systematic bias: incorrect well depth and incorrect center position.  Giving the GA a direct path to correct these systematic errors drastically speeds up convergence.

Selection is the final genetic operator and it is the process of evaluating the genes at the end of each generation to pick the best to remain in the population. At the start of selection, the fitness of all genes in the pool is calculated by:
\begin{multline}
\bar{\chi}_{E_{\rm binding}}^2 = \sum_{i} 0.1 \frac{(E_{i, {\rm exp}} - E_{i, {\rm pot}})^2}{\sigma_i^2} +\\
0.9 \frac{(\Delta E_{i, {\rm exp}} - \Delta E_{i, {\rm pot}})^2}{\sigma_i^2+\sigma_{i-1}^2},\\
\bar{\chi}_{B_v}^2 = \sum_{i} 0.1 \frac{({B_v}_{\ i, {\rm exp}} - {B_v}_{\ i, {\rm pot}})^2}{\sigma_i^2} +\\
0.9 \frac{(\Delta {B_v}_{\ i, {\rm exp}} - \Delta {B_v}_{\ i, {\rm pot}})^2}{\sigma_i^2+\sigma_{i-1}^2},
\end{multline}
\begin{equation}
\bar{\chi}_{\rm total}^2 = \bar{\chi}_{E_{\rm binding}}^2 + \bar{\chi}_{B_v}^2,
\end{equation}
where $E_{i, {\rm exp}}$ is the experimental binding energy of the i$^{\rm th}$ vibrational level with measured rotational constant ${B_v}_{\ i, {\rm exp}}$ and corresponding experimental uncertainties $\sigma_i$~\cite{footnote4}; $\Delta E_{i, {\rm exp}}$ is the experimental difference in binding energy between the i$^{\rm th}$ vibrational level and the i-1$^{\rm th}$ vibrational level; ${\Delta B_v}_{\ i, {\rm exp}}$ is the difference in rotational constants between the i$^{\rm th}$ vibrational level and the i-1$^{\rm th}$ vibrational level; and the quantities with `pot' subscripts denote those same quantities calculated from the trial potential. Because different aspects of the potential affect $E_{\rm binding}$ and $B_v$, we had the most success considering the errors independent of each other and as a multi-objective problem~\cite{deb2002fast}.  Additionally, we primarily use the energy spacing between vibrational levels, $\Delta E_{i}$, because the binding energy, $E_{i, {\rm binding}}$, depends on the binding energies of all previous vibrational levels while the nearest couple of points set the slope which governs $\Delta E_{i}$. This key insight empowers the crossover operator to swap small pieces of the potential that are `good' without negatively impacting higher vibrational levels. We added in a small fraction to our $\bar{\chi}^2$ calculation, with weighting factor 0.1, that pays attention to absolute binding energies to prevent large drifts in binding energies from accumulating in deep potential wells. We mirrored our $\bar{\chi}^2$ calculation of binding energy for rotational constants.  This unorthodox $\bar{\chi}^2$ calculation method prevents the genetic algorithm from getting stuck in sub-optimal solutions by reducing correlation between observables and it better reflects the raw experimental data.  More detail on our method can be found in the Appendix.

\begin{table}[h]
	\begin{tabular}{cccccc}
		\hline \hline
		r (\AA) & V(r) (cm$^{-1}$) & r (\AA) & V(r) (cm$^{-1}$) & r (\AA) & V(r) (cm$^{-1}$) \\
		\midrule[1.5pt]
		2.48 &	109.1142	& 4.1 &	-4909.8083	&	5.8	&	-1088.1540	\\
		2.52 &	-387.0396	& 4.2 &	-4649.2850	&	5.9	&	-970.6281	\\
		2.58 &	-1235.8825	& 4.3 &	-4378.3134	&	6	&	-863.8714	\\
		2.62 &	-1724.1180	& 4.4 &	-4101.8509	&	6.1	&	-769.5878	\\
		2.74 &	-3017.2003	& 4.5 &	-3824.2882	&	6.2	&	-683.1381	\\
		2.82 &	-3734.7812	& 4.6 &	-3549.4003	&	6.3	&	-608.4698	\\
		2.9 &	-4324.0701	& 4.7 &	-3280.1553	&	6.4	&	-539.5967	\\
		3.1 &	-5346.6398	& 4.8 &	-3018.8534	&	6.5	&	-480.5928	\\
		3.2 &	-5641.6278	& 4.9 &	-2767.0469	&	6.7	&	-379.3285	\\
		3.3 &	-5819.3234	& 5	&	-2526.0587	&	6.9	&	-301.2046	\\
		3.4 &	-5913.7376	& 5.1 &	-2297.4254	&	7.1	&	-239.8317	\\
		3.5 &	-5922.8906	& 5.2 &	-2081.9528	&	7.3	&	-190.6756	\\
		3.6 &	-5864.7259	& 5.3 &	-1880.2487	&	7.7	&	-125.2077	\\
		3.7 &	-5748.8936	& 5.4 &	-1693.5258	&	8.1	&	-79.7447	\\
		3.8 &	-5580.9393	& 5.5 &	-1521.0939	&	8.7	&	-51.8026	\\
		3.9 &	-5380.2337	& 5.6 &	-1362.9455	&	9.5	&	-28.9225	\\
		4.0 &	-5155.7034	& 5.7 &	-1219.2453	&	10.3 &	-16.4226	\\
		&		&		&		&	10.7	&	-7.1001	\\
		\hline
		$C_6$ & 12020368.0 &&  cm$^{-1}$ \AA$^6$ &&\\
		\hline \hline
	\end{tabular}
	\caption{The LiRb $X \ ^1\Sigma^+$ potential calculated by the genetic algorithm to 1\% accuracy in $\bar{\chi}^2$.  This potential is most accurate for low rotational states.  All energies reference the $^7$Li 2S$_{1/2}$ $F=2$ + $^{85}$Rb 5S$_{1/2}$ $F=2$ atomic asymptote.}
	\label{tab:Xpotential}
\end{table}

\section{$X \ ^1\Sigma^+ Potential$}

\begin{table} [b]
	\begin{tabular}{cccc}
		\hline \hline
		& Error {\it ab initio} & Error IPA & Error GA \\ \midrule[1.5pt]
		$\bar{\chi}^2_{E_{\rm binding}}$ & $9.01 \times 10^8$ & $1.95 \times 10^4$ & 18.59 \\
		$\bar{\chi}^2_{B_v}$ & $9.33 \times 10^3$ & 88.24 & 28.69 \\
		$\bar{\chi}^2_{\rm Total}$ & $9.01 \times 10^8$ & $1.96 \times 10^4$ & 47.28\\
		rms total (cm$^{-1}$) & 600.33 & 2.80 & 0.09 \\ \hline \hline
	\end{tabular} 
	\caption{Comparison of the errors in the different $X \ ^1\Sigma^+$ potentials.  The \textit{ab initio} potential is from Ref.~\cite{korek2009theoretical}, the IPA potential is from Ref.~\cite{ivanova2011x}, and Table~\ref{tab:Xpotential} presents the GA potential.  While the IPA potential matches the experimental data for high rotational states~\cite{ivanova2011x}, performance for low rotational states (reflected here) is not as good.}
	\label{tab:errorComparisonSmall}
\end{table}

\begin{figure}[t]
	\includegraphics[width=8.6 cm]{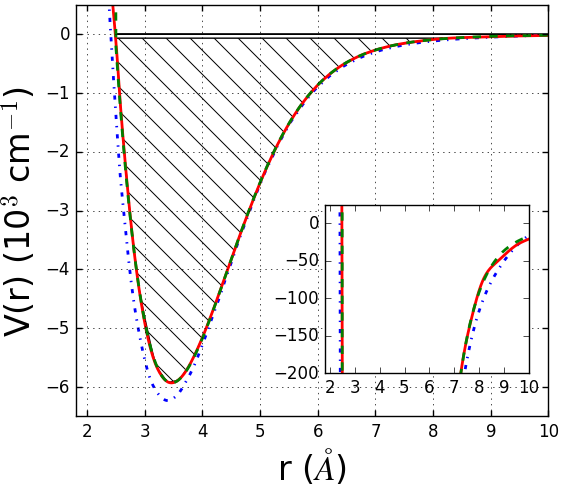}\\
	\caption{Comparison of the three $X \ ^1\Sigma^+$ potentials.  \textit{Ab initio} potential from Ref.~\cite{korek2009theoretical} is the blue dot-dashed curve; GA potential from Table~\ref{tab:Xpotential} is the red solid curve; IPA potential from Ref.~\cite{ivanova2011x} is the green dashed curve.  On the gross scale the GA and IPA potentials are identical.  The GA solver started with the \textit{ab initio} potential and improved it to its current state.  Because the \textit{ab initio} potential missed both the well depth and center position, this is a hard problem to solve.  The lower hashed area illustrates the region covered by the experimental data from Ref.~\cite{ivanova2011x}, while the top sliver shows the data from Ref.~\cite{dutta2017two}.  \textbf{Inset:} Comparison between the three potentials on an expanded scale using the same color scheme as before.}
	\label{fig:comparison}
\end{figure}

We chose to fit the $X \ ^1\Sigma^+$ state of LiRb to demonstrate our GA technique and we present our result in Table~\ref{tab:Xpotential}.  A rough sketch of our technique is as follows. We start with an \textit{ab initio} potential and some experimental data and calculate the vibrational energy $E_{\rm binding}$ and rotational constant $B_v$ for each vibrational level based on available experimental data. The genetic algorithm will descend to a solution of desired accuracy with its crossover, mutation and selection operators, calculating $E_{\rm binding}$ and $B_v$ for each of its trial potentials~\cite{le2016level} and comparing them to the experimental data.  We chose the $X \ ^1\Sigma^+$ state of LiRb state for a couple of reasons.  Highly accurate experimental data are available for most of the potential, $v=0-45$ and $50-51$~\cite{ivanova2011x,dutta2011laser,dutta2017two,footnote2}, and an experimental potential from Ivanova \textit{et al}~\cite{ivanova2011x} is available for comparison, fitted using the IPA method. Additionally, the \textit{ab initio} potential, while good, overestimates the well depth by about 400 cm$^{-1}$ and its equilibrium position is off by about 0.2 \AA.  These differences, combined with high accuracy experimental data make it a difficult fitting problem and a good testing ground.  The high accuracy data makes this problem more difficult because it requires a higher precision on the fitted points in the point-wise potential.  If our technique can fit the $X \ ^1\Sigma^+$ potential of LiRb, it can work for any potential.

Our algorithm reproduces the relevant rovibrational states for ultracold precision spectroscopy by fitting the bound state energies with total error less than 0.02 cm$^{-1}$ for $J<6$.  To generate our `guess' potential from the \textit{ab initio} potential of Ref.~\cite{korek2009theoretical}, we used a cubic spline function to connect the \textit{ab initio} points and picked an $r$-space grid arbitrarily. It took a couple of iterations of adding and subtracting points to our grid to converge.  There is nothing special about the \textit{ab initio} potential; we also reproduced the experimental data starting from a Morse potential. Details may be found in the appendix.  We derived $E_{\rm binding}$ and $B_v$ from the Dunham coefficients presented in Ref.~\cite{ivanova2011x} for $v=0-44$ and chose $\sigma_{E_{\rm binding}} = 0.02$ cm$^{-1}$ and $\sigma_{B_v} = 0.0003$ cm$^{-1}$ to achieve our desired uncertainty. The true experimental uncertainty in $B_v$ is usually closer to $0.0001$ cm$^{-1}$, but to do a more accurate fit, we would also need to include higher order rotational terms.  The experimental uncertainty in the energy difference between vibrational levels is usually about 0.005 cm$^{-1}$ although the absolute uncertainty in binding energy is 0.03 cm$^{-1}$~\cite{stevenson2016direct}.  We used the 2-photon data from Ref.~\cite{dutta2017two} for $v=50$ and $51$. For this region, the experimental uncertainties were $\sigma_{E_{\rm binding}} = 0.02$ cm$^{-1}$ and $\sigma_{B_v} = 0.005$ cm$^{-1}$.  Table~\ref{tab:errorComparisonSmall} shows the errors in our final potential, the initial \textit{ab initio} potential and the IPA potential from Ref.\cite{ivanova2011x} for easy comparison.  We were surprised by the large disagreement between our result and the IPA potential.  Upon further investigation, the experimental data used to derive the Dunham coefficients included high rotational states where higher order rotational terms are important.  The IPA potential matches the raw experimental data, but disagrees with the Dunham coefficients for low rotational states; likewise, we expect our potential to agree with the Dunham coefficients for low rotational states, but disagree for high rotational states.  The reader should note that the mis-match in datasets between our potential and the IPA potential prevents a direct comparison of the two methods; a full comparison is beyond the scope of this work.

\section{$C \ ^1\Sigma^+$ Potential}

\begin{figure}
	\includegraphics[width=8.6 cm]{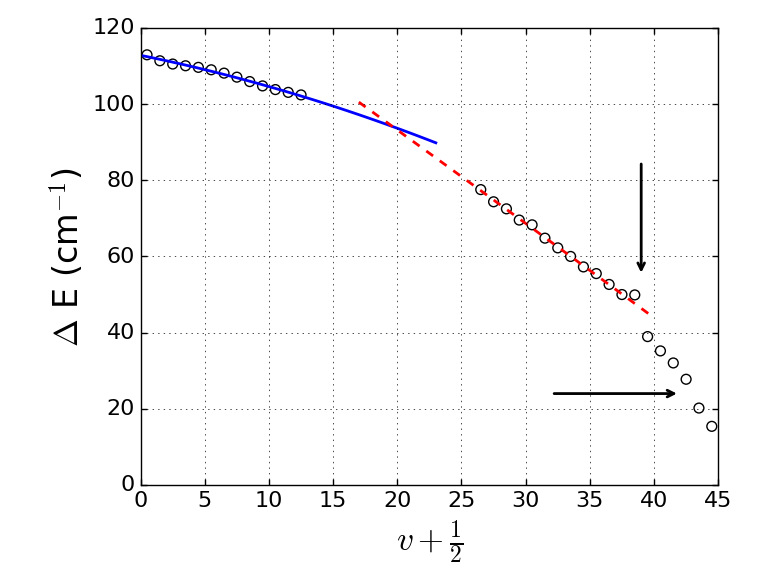}\\
	\caption{Vibrational spacing, $\Delta E = E_v - E_{v-1}$, for the $C \ ^1\Sigma^+$ state.  Black circles are experimental data from Ref.~\cite{ivanova2013b1pi,stevenson2016A}; blue dot-dashed line is the extrapolation from low vibrational levels into the gap in experimental data; red dashed line is extrapolation from high vibrational levels; black arrows highlight two jumps in the data. We extrapolated from the observed vibrational levels to estimate the binding energy of $v=14-25$. The vertical arrow highlights $v=38$ and $39$ and the horizontal arrow highlights $v=44$ and $45$. We omitted these vibrational levels from the fit because they caused an non-physical potential.}
	\label{fig:C_data}
\end{figure}

The $C \ ^1\Sigma^+$ state of LiRb also has highly accurate experimental data for both rotational constants and vibrational binding energies.  The $C \ ^1\Sigma^+, \ v=0-11$ levels, with approximate uncertainties $\sigma_{E_{\rm binding}} = 0.02$ cm$^{-1}$ and $\sigma_{B_v} = 0.0001$ cm$^{-1}$, were originally observed through perturbations on the $B \ ^1\Pi$ state~\cite{ivanova2013b1pi,dutta2011laser}.  More recently resonantly-enhanced-two-color-ionization (RE2PI)~\cite{stevenson2016A} measured $v=$12, 13 and $26-45$ with approximate uncertainty $\sigma_{E_{\rm binding}} = 0.3$ cm$^{-1}$, but unfortunately RE2PI is not sensitive enough to measure rotational structure. To fill in the gap in observed vibrational levels, we used the molecular constants extracted from $v=0-13$ and $v=26-40$ to estimate the binding energies for $v=14-25$, as shown in Fig.~\ref{fig:C_data}.  We used the extrapolation from low vibrational levels for estimating $v=14-21$ and the extrapolation from high vibrational levels for estimating $v=22-25$.  This lead to projected uncertainties between $0.5$ and $1.6$ cm$^{-1}$ for $v=14-25$.

\begin{table}
	\begin{tabular}{cccc}
		\hline \hline
		& Error {\it ab initio} & Error IPA & Error GA \\ \midrule[1.5pt]
		$\bar{\chi}^2_{E_{\rm binding}}$ & $2.53 \times 10^7$ & $1.60 \times 10^5$ & 61.37 \\
		$\bar{\chi}^2_{B_v}$ & $6.87 \times 10^3$ & 139.41 & 1.17 \\
		$\bar{\chi}^2_{\rm Total}$ & $2.53 \times 10^7$ & $1.60 \times 10^5$ & 62.54\\
		rms total (cm$^{-1}$) & $2.51 \times 10^3$ & $1.26 \times 10^3$ & 36.25 \\ \hline \hline
	\end{tabular} 
	\caption{Comparison of the error in the different $C \ ^1\Sigma^+$ potentials.  The total rms error is much larger than for the $X \ ^1\Sigma^+$ state because of lower accuracy experimental data and unobserved vibrational levels.  We estimated the locations of the unobserved levels with the molecular constants, resulting in high uncertainty.  The large error in the IPA potential is because there wasn't data available for the higher vibrational levels at the time it was constructed.}
	\label{tab:errorC}
\end{table}

\begin{table} 
	\begin{tabular}{ccccccc}
		\hline \hline
		r (\AA) & V(r) (cm$^{-1}$) & r (\AA) & V(r) (cm$^{-1}$) & r (\AA) & V(r) (cm$^{-1}$) \\
		\midrule[1.5pt]
		3 & 385.0443 & 5 & -3133.6171 & 7.2 & -640.0546 \\
		3.12 & -676.4168 & 5.2 & -2888.4813 & 7.4 & -525.0763 \\
		3.24 & -1448.7373 & 5.4 & -2625.8863 & 7.6 & -428.3892 \\
		3.4 & -2105.8284 & 5.6 & -2350.4856 & 7.8 & -332.5077 \\
		3.6 & -2788.3000 & 5.8 & -2057.1814 & 8 & -266.3663 \\
		3.8 & -3235.5227 & 6 & -1783.2375 & 8.3 & -167.1230 \\
		4 & -3482.6914 & 6.2 & -1516.6927 & 8.6 & -111.1909 \\
		4.2 & -3592.0639 & 6.4 & -1289.0574 & 8.9 & -76.7587 \\
		4.4 & -3584.9191 & 6.6 & -1112.6115 & 9.2 & -20.9949 \\
		4.6 & -3494.7808 & 6.8 & -929.7609 & 9.8 & 19.6301 \\
		4.8 & -3340.4983 & 7 & -777.4679 & 10.4 & 38.9005 \\
		\hline
		$C_6$ & -50420955.0 &&cm$^{-1}$ \AA$^6$&&\\
		\hline \hline
	\end{tabular}
	\caption{The $C \ ^1\Sigma^+$ state of LiRb calculated by the genetic algorithm to 1 \% accuracy as measured by $\bar{\chi}^2$.  This potential is most accurate for low rotational states. All energies reference the Li 2P$_{1/2}$ + Rb 5S$_{1/2}$ atomic asymptote.}
	\label{tab:Cpotential}
\end{table}

\begin{figure} [t]
	\includegraphics[width=8.6 cm]{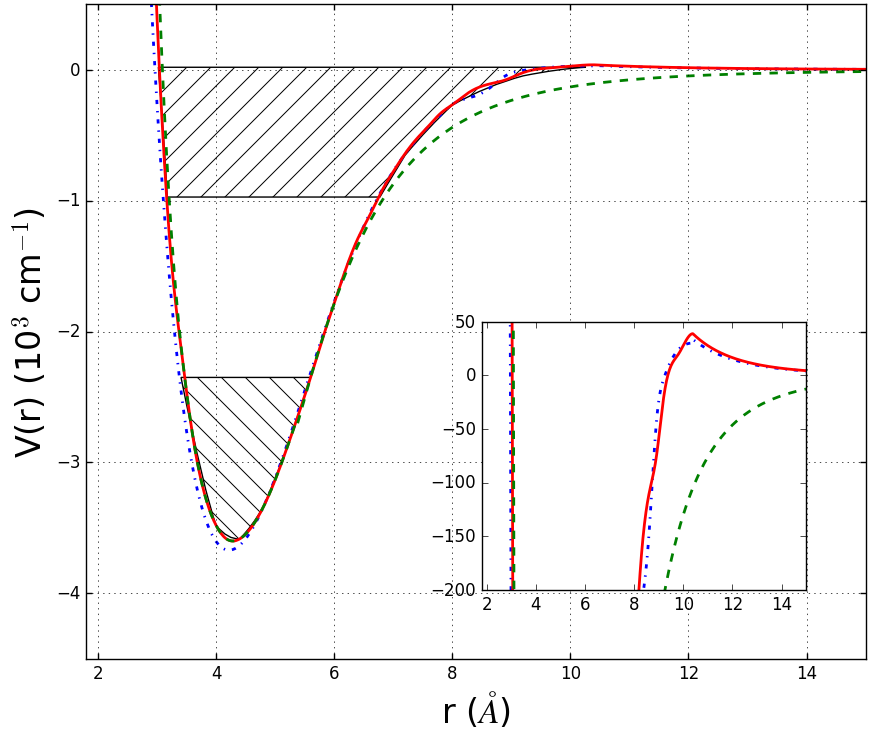}\\
	\caption{Comparison of $C \ ^1\Sigma^+$ potentials. \textit{Ab initio} potential from Ref.~\cite{korek2009theoretical} is the blue dot-dashed line; IPA potential from Ref.~\cite{ivanova2013b1pi} is the green dashed line; GA potential from Table.~\ref{tab:Cpotential} is the red solid line.  The lower hashed region fits the experimental data from Ref.~\cite{ivanova2013b1pi} and the upper hashed region fits data from Ref.~\cite{stevenson2016A}.  \textbf{Inset:} a close up comparison of the same potentials with the same color scheme showing the differences at low energy.  Irregularities in the experimental data cause the unevenness of the GA potential, discussed in the text and illustrated in Fig.~\ref{fig:C_data}.}
	\label{fig:C1Sigma}
\end{figure}

For the rotational constants, data is available for $v<12$ because of perturbations these levels caused to the $B \ ^1\Pi$ spectra~\cite{ivanova2013b1pi}.  After studying the \textit{ab initio} potentials for the $X \ ^1\Sigma^+$, $B \ ^1\Pi$ and $C \ ^1\Sigma^+$ states, and comparing them to the experimental data, we concluded that there was a systematic shift in the calculated rotational constants, but that once the perturbation was properly accounted for, the rotational constants were accurate.  For example, for $v=0$ and $1$ of the $C \ ^1\Sigma^+$ state, the \textit{ab initio} rotational constants were 0.146 and 0.145 cm$^{-1}$, while they were measured to be 0.141 and 0.140 cm$^{-1}$.  Notice how, although both constants are off, and they are off by the same amount, 0.005 cm$^{-1}$.  For the states with data on rotational spectra, we set the uncertainty to 0.0001 cm$^{-1}$ and then used the $B_v$ progression, paired with the \textit{ab initio} potential to extrapolate the remaining vibrational levels.  To account for the inaccuracies this will introduce, we set the uncertainty to 0.005 cm$^{-1}$ for $v=12-39$, 0.01 cm$^{-1}$ for $v=40-42$ and 0.02 cm$^{-1}$ for $v>42$.  In total, the $\bar{\chi}^2$ and rms error for our fitted $C \ ^1\Sigma^+$ potential are given in Table~\ref{tab:errorC}, compared to the \textit{ab initio} potential~\cite{korek2009theoretical} and the IPA potential from Ref.~\cite{ivanova2013b1pi}.  Table~\ref{tab:Cpotential} presents the fitted $C \ ^1\Sigma^+$ potential, also plotted in Fig.~\ref{fig:C1Sigma}.

There were four experimental data points omitted from the fit.  Close inspection of the experimental data, shown in Fig.~\ref{fig:C_data}, reveals that the vibrational progression has two hard jumps between $v=38$ and $v=39$ and between $v=43$ and $v=44$.  We suspect a mis-assignment of the experimental data caused the hard jump between $v=38$ to $v=39$.  As seen in Fig.~[2] of Ref.~\cite{stevenson2016A}, the RE2PI progression is very hard to pick out for these vibrational levels.  As such $v=38-39$ have been omitted from the fit.  As seen in Fig.~[3] of Ref.~\cite{stevenson2016A}, the progression is quite clear for $v=44$ and $v=45$, but we think a perturbation between $C \ ^1\Sigma^+ \ v=44$ and $D \ ^1\Pi \ v=20$  causes the unevenness.  Ultimately, we omitted $v=44$ and $45$ because they resulted in a non-physical potential.

\section{Conclusion}

In conclusion, we present a novel technique for fitting the potential energy curves of diatomic molecules. We used the high accuracy data for the $X \ ^1\Sigma^+$ state of lithium-rubidium to benchmark our technique, where it was able to out-perform the state-of-the-art potentials for low rotational states (of particular relevance for ultracold applications).  We combined our technique and recent experimental data to improve the available potential for the $C \ ^1\Sigma^+$ state of lithium-rubidium. We are working on future improvements to the present technique to deal with strong spin-orbit coupled systems, where the current state of the art struggles, such as simultaneously fitting the complicated $A-b$ complex that occurs in the bi-alkali family. Additionally, we are looking to improve our technique following in the path laid out by Price and Storn~\cite{price2006differential,storn1997differential} in moving from an annealing based GA to differential evolution. We hope that because the differential evolution heuristic is quite simple, other experimental bi-alkali teams will be able to easily implement our method. Finally, owing to the nature of GA's, and its straight forward multidimensional generalization, we believe that this approach may be suitable for fitting multidimensional potential energy surfaces, provided a reasonable number of points can describe the interaction.

\section{Acknowledgements}

We would like to thank Sourav Dutta for conversations that started us down this path, Roman Krems for his work in paving the way for machine learning techniques in cold chemistry and helpful suggestions, and Dan Elliott for reviewing early manuscripts.  We acknowledge support from Purdue University and Le Bleu.


\begin{thebibliography}{37}
\bibitem[{foo({\natexlab{a}})}]{footnote5}
	\bibinfo{note}{We found the region $r<r_{in}$ has little impact on our
		observables}.
\bibitem[{foo({\natexlab{b}})}]{footnote4}
	\bibinfo{note}{$\sigma_i$ need not be the same for $E_{\rm binding}$ and
		$B_v$}.
\bibitem[{foo({\natexlab{c}})}]{footnote2}
	\bibinfo{note}{$v=45$ was omitted from the fit because it resulted in a
		non-physical potential. More experimental data on $v=45-50$ is needed to
		understand why.}
\end{thebibliography}


\appendix

\section{Method Details}

\subsection{Pseudo-code}

\begin{figure}[b]
	\includegraphics[width=8.6cm]{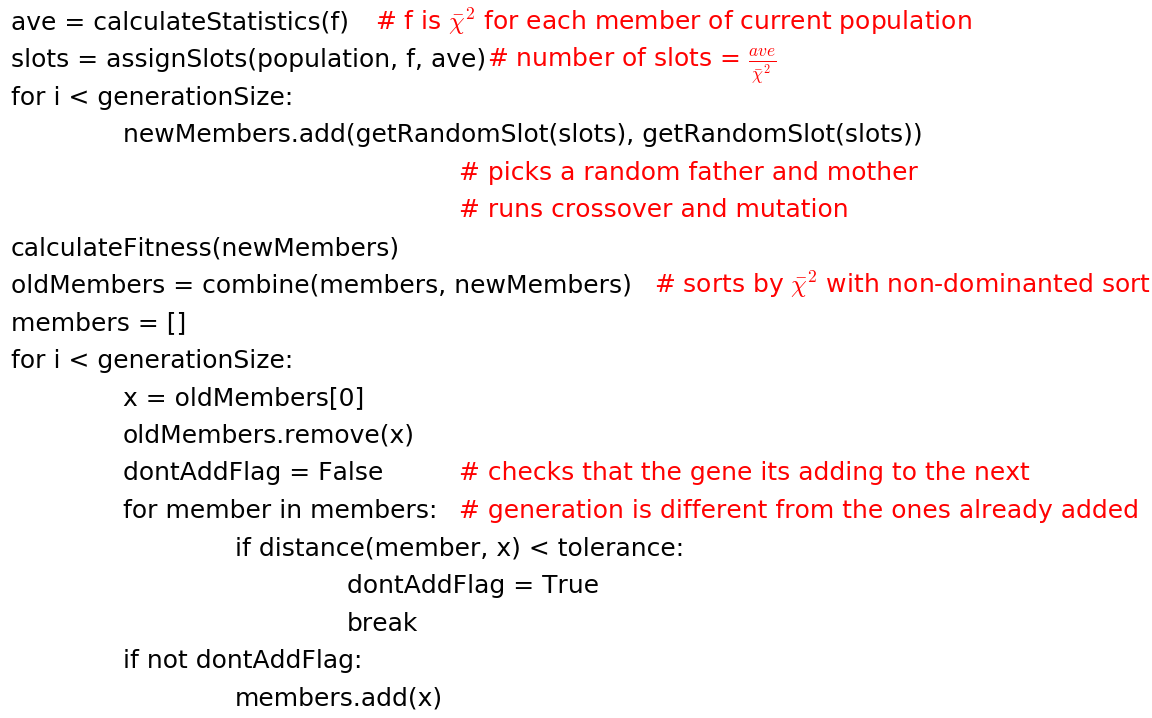}
	\caption{The main loop of the GA in pseudo code.}
	\label{fig:mainLoop}
\end{figure}

The pseudo-code for the main loop of the genetic algorithm is presented in Fig.~\ref{fig:mainLoop}.  At the start of the generation, the algorithm calculates $\bar{\chi}^2$ for each gene and then compiles the average for each minimization objective (binding energy and rotational constants).  It uses this information to calculate the value of each gene using the roulette method.  For example, if the gene has $\bar{\chi}_{\rm total}^2 = (\bar{\chi}_{\rm B_v}^2, \bar{\chi}_{E_{\rm binding}}^2) = (100, 45)$ and the average is $(200, 50)$, its value is $mean(\frac{(200, 50)}{(100, 45)}) = mean(2,1.11) = 1.56$.  The total value of the population is computed and then the probability of each gene being picked for either crossover or mutation is the fraction of total value it represents.  In the prior example, if the total value in the population is 50, then the example gene has a 0.0312 chance to be picked for each new gene generated by crossover or mutation.

After the new genes are formed, their fitness is also evaluated and then they are combined with the fast non-dominated sort from NSGA-II~\cite{deb2002fast}.  From here, the new generation is formed by taking the best 100 genes, provided that no two genes are closer than some small distance $t \approx 0.1$ apart.  This prevents identical good but not prefect solutions from running away too early in the descent.  Eventually, the best solution available will reach some threshold, usually around $\bar{\chi}^2_{\rm total} = 500$, at which point the genetic algorithm will call a Levenberg-Marquardt based gradient solver to reach full convergence.  This simply helps speed up convergence once the genetic algorithm is sufficiently close to a true solution (i.e. it has found the correct minima in the N-dimensional space in which it is searching).

\subsection{Crossover}

We tried several variants of crossover, such as two point or orthogonal, but found that the more sophisticated algorithms worked worse than the simple one point method.  Ultimately, we used a blended one point crossover in our genetic algorithm:
\begin{multline}
	\vec{x} = (v_0, . . . v_{i-4}, v_{i-3}, v_{i-2}, v_{i-1}, v_{i}, v_{i+1}, v_{i+2}, v_{i+3}, v_{i+4}, . . . v_N) \\ \times (1, . . . 1, \frac{7}{8}, \frac{6}{8}, \frac{5}{8}, \frac{4}{8}, \frac{3}{8}, \frac{2}{8}, \frac{1}{8}, 0, . . . 0) + \vec{u} \times \vec{u}_{mask}, 
	\label{Eq:crossover}
\end{multline}
where $\vec{x}$ is the new gene created during crossover, $\vec{v}$ and $\vec{u}$ are the original genes, $\vec{v}_{mask}$ is illustrated and $\vec{u}_{mask} = \vec{1} - \vec{v}_{mask}$.  In effect the blending averages around the crossover point which prevents hard jumps from forming in the potentials and leading to non-physical results.  This helps the crossover operator stay in the stable and physical solution space for the potentials when it uses two dissimilar potentials.

\subsection{Mutation}

We changed the probability and amplitude (known as annealing) of all of our mutation operators as the algorithm descended to a solution.  We changed the variance of our one point mutation operator, 
\begin{equation}
	V(r_i) = V_{\rm old}(r_i)(1+x),
	\label{Eq:mutation}
\end{equation}
where $x$ is a normally distributed random variable with 0 mean and variance $\sigma^2$, according to
\begin{equation}
	\sigma = 
	\begin{cases}
		0.02 & \bar{\chi}^2_{\rm total} \geq 10^5 \\
		2 \times \bar{\chi}^2_{\rm total} \times 10^{-7}& 10^5 \geq \bar{\chi}^2_{\rm total} \geq 25 \\
		5 \times 10^{-6} & 25 \geq \bar{\chi}^2_{\rm total}
	\end{cases},
\end{equation}
where $\bar{\chi}^2$ is the total error in the potential being mutated.  The different numbers, although empirically chosen, do follow some logic.  The largest changes this mutation operator can make and keep a stable potential are about 2 \% which is the upper bound.  The precision we are striving for, 0.02 cm$^{-1}$, is about $5 \times 10^{-6}$ smaller than the absolute value of the largest data point, approximately 6000 cm$^{-1}$.  The $\bar{\chi}^2_{\rm total}$ value where the annealing starts, $10^5$, was picked because this was when the genetic algorithm started to stall when modifying points by 2\%.  And finally the linear region simply connects the two extremes systematically.  All of our mutation operators follow a similar annealing procedure, although with different values picked in a similar fashion.  Additionally, the probabilities of the different mutation operators were changed throughout.  For example, our well depth mutation operator has the probability:
\begin{equation}
	P = 
	\begin{cases}
		0.3 & \bar{\chi}_{vib}^2 \geq 10^8 \\
		0.3 (1 - \bar{\chi}_{vib}^2 \times 10^{-8}) & 10^8 \geq \bar{\chi}_{vib}^2 \geq 10^7 \\
		0.01 & 10^7 \geq \bar{\chi}_{vib}^2
	\end{cases},
\end{equation}
where $\bar{\chi}_{vib}^2$ is the error only in the binding energies.  Again the number are empirical and they follow a similar logic to before.  The three macro-scale mutation operators, changing well depth or either wall position have a high probability at the start and rapidly drop off as the algorithm converges. Naturally, the one point mutation operator is the converse and by the end is the dominant mutation operator.

\section{Program Specs}

To generate our `guess' potential from the \textit{ab initio} potential of Ref.~\cite{korek2009theoretical} and we gave the spline form of the \textit{ab initio} potential to the genetic algorithm which generated slight variants with its mutation operators for an initial population of 100 genes.  The genetic algorithm ran on a single core of an i-3 processor; it took around 500 generations to converge to 1 \% uncertainty in a little over 30 minutes.  In total it called the Numerov solver to find the bound states roughly $5 \times 10^4$ times.  Although a 30 minute run time is not excessive, it could certainly be improved upon: running on a better processor is the easiest improvement; this algorithm lends itself nicely to parallelization and could result in about a 4x increase in speed; our code is a Fortran-Python hybrid and translating it entirely into Fortran could result in modest gains in speed.  Finally, there are a few additional variables that need to be specified for any genetic algorithm.  We used a population size of 100, and generated 40 new genes using crossover and 40 using mutation each generation.

\section{Morse Potential}

We also fitted the $X \ ^1\Sigma^+$ state of LiRb starting from a Morse potential to demonstrate the robustness of our fitting routine.  Morse potentials are written as:
\begin{equation}
	V(r) = D_e (e^{-2a(r-r_e)}-2e^{-a(r-r_e)}),
\end{equation}
where $D_e$ is the dissociation energy of the potential, $r_e$ is the equilibrium position of the molecule in the $v = 0$ state, and $a$ is a constant related to the width of the potential.  We first tried a Morse potential defined by $D_e = 6000$ cm$^{-1}$, $r_e = 3.45$ \AA, and $a = 0.78$ \AA, which our routine fitted in 600 generations, a comparable run time to when it started with the \textit{ab initio} potential.  We also tried a much worse starting potential, where we made $D_e = 3000$ cm$^{-1}$ but kept the other parameters the same.  The program was able to converge from this starting point in 1140 generations.

\section{Comparison Tables}

In the two tables in this section we provide a full comparison between our results, the \textit{ab initio} potential~\cite{korek2009theoretical} and the IPA potential~\cite{ivanova2011x,ivanova2013b1pi} for every vibrational level with experimental data.

\begin{table*}
	\begin{tabular}{ccccccccc}
		\hline \hline
		$v$ & $E_{binding}$ Exp & $E_{binding}$ \textit{ab initio}& $E_{binding}$ GA& $E_{binding}$ IPA& $B_v$ Exp & $B_v$ \textit{ab initio} & $B_v$ GA & $B_v$ IPA \\ \midrule[1.5pt]
		0 & -5830.84 (0.02) & -6129.58 & -5830.84 & -5830.83 & 0.216 (0.0003) & 0.225 & 0.216 & 0.216 \\
		1 & -5637.86 (0.02) & -5936.34 & -5637.84 & -5637.85 & 0.214 (0.0003) & 0.224 & 0.215 & 0.214 \\
		2 & -5447.13 (0.02) & -5748.0 & -5447.12 & -5447.13 & 0.213 (0.0003) & 0.223 & 0.214 & 0.213 \\
		3 & -5258.67 (0.02) & -5561.78 & -5258.66 & -5258.67 & 0.211 (0.0003) & 0.22 & 0.212 & 0.211 \\
		4 & -5072.49 (0.02) & -5375.93 & -5072.48 & -5072.5 & 0.209 (0.0003) & 0.219 & 0.21 & 0.21 \\
		5 & -4888.61 (0.02) & -5190.06 & -4888.6 & -4888.64 & 0.208 (0.0003) & 0.218 & 0.208 & 0.208 \\
		6 & -4707.05 (0.02) & -5007.32 & -4707.05 & -4707.1 & 0.206 (0.0003) & 0.216 & 0.206 & 0.206 \\
		7 & -4527.84 (0.02) & -4827.21 & -4527.81 & -4527.91 & 0.205 (0.0003) & 0.215 & 0.205 & 0.205 \\
		8 & -4351.0 (0.02) & -4649.7 & -4351.0 & -4351.11 & 0.203 (0.0003) & 0.213 & 0.203 & 0.203 \\
		9 & -4176.56 (0.02) & -4475.04 & -4176.55 & -4176.71 & 0.201 (0.0003) & 0.211 & 0.201 & 0.201 \\
		10 & -4004.54 (0.02) & -4303.28 & -4004.54 & -4004.74 & 0.199 (0.0003) & 0.209 & 0.199 & 0.2 \\
		11 & -3834.99 (0.02) & -4132.45 & -3834.99 & -3835.24 & 0.198 (0.0003) & 0.208 & 0.197 & 0.198 \\
		12 & -3667.94 (0.02) & -3962.15 & -3667.94 & -3668.25 & 0.196 (0.0003) & 0.206 & 0.196 & 0.196 \\
		13 & -3503.42 (0.02) & -3794.77 & -3503.42 & -3503.8 & 0.194 (0.0003) & 0.203 & 0.194 & 0.194 \\
		14 & -3341.47 (0.02) & -3630.38 & -3341.48 & -3341.94 & 0.192 (0.0003) & 0.201 & 0.192 & 0.192 \\
		15 & -3182.15 (0.02) & -3467.06 & -3182.15 & -3182.7 & 0.19 (0.0003) & 0.2 & 0.19 & 0.19 \\
		16 & -3025.5 (0.02) & -3306.31 & -3025.49 & -3026.13 & 0.188 (0.0003) & 0.197 & 0.188 & 0.188 \\
		17 & -2871.57 (0.02) & -3150.22 & -2871.56 & -2872.3 & 0.186 (0.0003) & 0.194 & 0.185 & 0.186 \\
		18 & -2720.41 (0.02) & -2996.91 & -2720.41 & -2721.25 & 0.184 (0.0003) & 0.193 & 0.183 & 0.184 \\
		19 & -2572.1 (0.02) & -2844.74 & -2572.1 & -2573.04 & 0.182 (0.0003) & 0.192 & 0.181 & 0.182 \\
		20 & -2426.69 (0.02) & -2695.1 & -2426.68 & -2427.73 & 0.18 (0.0003) & 0.19 & 0.179 & 0.18 \\
		21 & -2284.25 (0.02) & -2548.78 & -2284.25 & -2285.41 & 0.177 (0.0003) & 0.187 & 0.177 & 0.177 \\
		22 & -2144.86 (0.02) & -2405.4 & -2144.87 & -2146.13 & 0.175 (0.0003) & 0.185 & 0.174 & 0.175 \\
		23 & -2008.61 (0.02) & -2265.12 & -2008.62 & -2009.98 & 0.172 (0.0003) & 0.182 & 0.172 & 0.173 \\
		24 & -1875.58 (0.02) & -2128.17 & -1875.58 & -1877.04 & 0.17 (0.0003) & 0.179 & 0.169 & 0.17 \\
		25 & -1745.87 (0.02) & -1994.65 & -1745.85 & -1747.39 & 0.167 (0.0003) & 0.176 & 0.167 & 0.167 \\
		26 & -1619.57 (0.02) & -1863.83 & -1619.56 & -1621.15 & 0.164 (0.0003) & 0.174 & 0.164 & 0.164 \\
		27 & -1496.8 (0.02) & -1735.06 & -1496.81 & -1498.39 & 0.161 (0.0003) & 0.171 & 0.161 & 0.161 \\
		28 & -1377.68 (0.02) & -1610.09 & -1377.69 & -1379.23 & 0.159 (0.0003) & 0.167 & 0.158 & 0.158 \\
		29 & -1262.33 (0.02) & -1489.54 & -1262.33 & -1263.8 & 0.155 (0.0003) & 0.164 & 0.155 & 0.155 \\
		30 & -1150.88 (0.02) & -1371.5 & -1150.9 & -1152.2 & 0.152 (0.0003) & 0.161 & 0.152 & 0.151 \\
		31 & -1043.49 (0.02) & -1257.69 & -1043.51 & -1044.59 & 0.148 (0.0003) & 0.156 & 0.148 & 0.148 \\
		32 & -940.3 (0.02) & -1148.96 & -940.3 & -941.13 & 0.145 (0.0003) & 0.153 & 0.144 & 0.144 \\
		33 & -841.48 (0.02) & -1042.44 & -841.49 & -841.97 & 0.141 (0.0003) & 0.151 & 0.141 & 0.14 \\
		34 & -747.21 (0.02) & -940.43 & -747.24 & -747.31 & 0.137 (0.0003) & 0.146 & 0.136 & 0.135 \\
		35 & -657.67 (0.02) & -843.64 & -657.68 & -657.34 & 0.133 (0.0003) & 0.142 & 0.132 & 0.131 \\
		36 & -573.07 (0.02) & -751.22 & -573.07 & -572.3 & 0.128 (0.0003) & 0.137 & 0.128 & 0.126 \\
		37 & -493.6 (0.02) & -663.89 & -493.57 & -492.4 & 0.123 (0.0003) & 0.133 & 0.123 & 0.121 \\
		38 & -419.48 (0.02) & -580.86 & -419.51 & -417.89 & 0.118 (0.0003) & 0.129 & 0.118 & 0.115 \\
		39 & -350.94 (0.02) & -503.42 & -350.93 & -349.01 & 0.112 (0.0003) & 0.123 & 0.112 & 0.11 \\
		40 & -288.2 (0.02) & -431.59 & -288.19 & -286.02 & 0.106 (0.0003) & 0.118 & 0.106 & 0.104 \\
		41 & -231.49 (0.02) & -364.94 & -231.53 & -229.15 & 0.1 (0.005) & 0.112 & 0.1 & 0.097 \\
		42 & -181.02 (0.02) & -303.75 & -181.02 & -178.64 & 0.093 (0.005) & 0.107 & 0.093 & 0.09 \\
		43 & -136.98 (0.02) & -248.62 & -136.97 & -134.68 & 0.085 (0.005) & 0.099 & 0.085 & 0.083 \\
		44 & -99.51 (0.02) & -199.35 & -99.5 & -97.4 & 0.078 (0.005) & 0.093 & 0.078 & 0.075 \\
		45 &  & -156.53 & -67.96 & -66.84 &  & 0.086 & 0.07 & 0.067 \\
		46 &  & -119.48 & -45.44 & -42.91 &  & 0.078 & 0.056 & 0.058 \\
		47 &  & -88.38 & -29.13 & -25.32 &  & 0.071 & 0.051 & 0.049 \\
		48 &  & -62.87 & -16.31 & -13.45 &  & 0.063 & 0.043 & 0.04 \\
		49 &  & -41.7 & -5.73 & -6.13 &  & 0.057 & 0.037 & 0.03 \\
		50 & -1.82 (0.02) & -23.95 & -1.78 & -2.2 &  & 0.051 & 0.021 & 0.021 \\
		51 & -0.41 (0.02) & -9.63 & -0.29 & -0.55 & 0.013 (0.01) & 0.044 & 0.012 & 0.013 \\
		\hline \hline
	\end{tabular}
	\caption{Full comparison of 3 potentials for the $X \ ^1\Sigma^+$ state in LiRb, all numbers are in cm$^{-1}$. Blank entries denote vibrational levels for which there is no exp data or there is a flaw in the exp data.  The number in parenthesis is the error in the experimental value.}
	\label{tab:Xcomparison}
\end{table*}

\begin{table*}
	\begin{tabular}{ccccccccc}
		\hline \hline
		$v$ & $E_{binding}$ Exp & $E_{binding}$ \textit{ab initio}& $E_{binding}$ GA& $E_{binding}$ IPA& $B_v$ Exp & $B_v$ \textit{ ab initio} & $B_v$ GA & $B_v$ IPA \\ \midrule[1.5pt]
		0 & -3544.05 (0.02) & -3614.51 & -3544.05 & -3544.21 & 0.141 (0.0001) & 0.146 & 0.141 & 0.141 \\
		1 & -3431.12 (0.02) & -3501.62 & -3431.1 & -3431.36 & 0.14 (0.0001) & 0.145 & 0.14 & 0.14 \\
		2 & -3319.75 (0.02) & -3389.44 & -3319.77 & -3319.91 & 0.139 (0.0001) & 0.144 & 0.139 & 0.139 \\
		3 & -3209.24 (0.02) & -3277.97 & -3209.27 & -3209.33 & 0.138 (0.0001) & 0.143 & 0.138 & 0.138 \\
		4 & -3099.18 (0.02) & -3167.25 & -3099.18 & -3099.23 & 0.137 (0.0001) & 0.142 & 0.137 & 0.137 \\
		5 & -2989.53 (0.02) & -3057.34 & -2989.54 & -2989.64 & 0.136 (0.0001) & 0.141 & 0.136 & 0.136 \\
		6 & -2880.53 (0.02) & -2948.3 & -2880.53 & -2880.84 & 0.135 (0.0001) & 0.14 & 0.135 & 0.135 \\
		7 & -2772.41 (0.02) & -2840.18 & -2772.4 & -2773.22 & 0.134 (0.0001) & 0.139 & 0.134 & 0.134 \\
		8 & -2665.34 (0.02) & -2733.0 & -2665.33 & -2667.45 & 0.133 (0.0001) & 0.138 & 0.133 & 0.132 \\
		9 & -2559.43 (0.02) & -2626.81 & -2559.39 & -2563.95 & 0.132 (0.0001) & 0.137 & 0.132 & 0.131 \\
		10 & -2454.66 (0.02) & -2521.67 & -2454.59 & -2461.1 & 0.131 (0.0001) & 0.136 & 0.131 & 0.131 \\
		11 & -2350.85 (0.02) & -2417.63 & -2350.86 & -2356.17 & 0.13 (0.0001) & 0.135 & 0.13 & 0.131 \\
		12 & -2247.78 (0.5) & -2314.73 & -2248.08 & -2250.03 & 0.129 (0.005) & 0.134 & 0.129 & 0.13 \\
		13 & -2145.38 (0.5) & -2213.04 & -2146.16 & -2146.7 & 0.128 (0.005) & 0.132 & 0.129 & 0.128 \\
		14 & -2045.22 (0.55) & -2112.6 & -2045.09 & -2045.84 & 0.127 (0.005) & 0.131 & 0.128 & 0.127 \\
		15 & -1946.47 (0.71) & -2013.47 & -1944.95 & -1944.51 & 0.125 (0.005) & 0.13 & 0.127 & 0.126 \\
		16 & -1849.17 (0.87) & -1915.71 & -1845.85 & -1846.06 & 0.124 (0.005) & 0.129 & 0.127 & 0.124 \\
		17 & -1753.41 (1.02) & -1819.38 & -1748.06 & -1749.62 & 0.123 (0.005) & 0.127 & 0.125 & 0.123 \\
		18 & -1659.24 (1.17) & -1724.55 & -1651.98 & -1654.2 & 0.121 (0.005) & 0.126 & 0.124 & 0.121 \\
		19 & -1566.74 (1.32) & -1631.27 & -1558.11 & -1562.17 & 0.12 (0.005) & 0.124 & 0.122 & 0.119 \\
		20 & -1475.97 (1.48) & -1539.63 & -1466.88 & -1471.04 & 0.118 (0.005) & 0.123 & 0.119 & 0.117 \\
		21 & -1386.98 (1.64) & -1449.68 & -1378.58 & -1383.46 & 0.117 (0.005) & 0.121 & 0.117 & 0.115 \\
		22 & -1301.01 (1.67) & -1361.51 & -1293.34 & -1297.16 & 0.115 (0.005) & 0.12 & 0.114 & 0.113 \\
		23 & -1215.8 (1.49) & -1275.2 & -1210.79 & -1214.35 & 0.113 (0.005) & 0.118 & 0.112 & 0.111 \\
		24 & -1132.57 (1.25) & -1190.83 & -1130.14 & -1133.31 & 0.111 (0.005) & 0.116 & 0.11 & 0.108 \\
		25 & -1051.37 (0.91) & -1108.5 & -1050.68 & -1055.44 & 0.109 (0.005) & 0.114 & 0.108 & 0.106 \\
		26 & -972.28 (0.2) & -1028.29 & -972.2 & -980.16 & 0.107 (0.005) & 0.112 & 0.107 & 0.103 \\
		27 & -894.73 (0.2) & -950.29 & -895.14 & -907.37 & 0.105 (0.005) & 0.11 & 0.105 & 0.101 \\
		28 & -820.38 (0.2) & -874.59 & -820.32 & -837.93 & 0.103 (0.005) & 0.108 & 0.103 & 0.098 \\
		29 & -747.88 (0.2) & -801.29 & -748.09 & -770.93 & 0.101 (0.005) & 0.106 & 0.101 & 0.095 \\
		30 & -678.33 (0.2) & -730.43 & -678.1 & -706.87 & 0.099 (0.005) & 0.104 & 0.099 & 0.093 \\
		31 & -610.03 (0.2) & -662.09 & -610.27 & -646.07 & 0.097 (0.005) & 0.101 & 0.096 & 0.09 \\
		32 & -545.23 (0.2) & -596.28 & -545.12 & -587.97 & 0.095 (0.005) & 0.099 & 0.094 & 0.087 \\
		33 & -482.98 (0.2) & -533.01 & -482.98 & -532.75 & 0.092 (0.005) & 0.096 & 0.091 & 0.084 \\
		34 & -422.98 (0.2) & -472.22 & -423.5 & -480.73 & 0.09 (0.005) & 0.094 & 0.089 & 0.082 \\
		35 & -365.73 (0.2) & -413.83 & -366.03 & -431.74 & 0.088 (0.005) & 0.091 & 0.087 & 0.078 \\
		36 & -310.23 (0.2) & -357.75 & -310.54 & -385.57 & 0.086 (0.005) & 0.089 & 0.085 & 0.075 \\
		37 & -257.58 (0.2) & -303.95 & -258.24 & -342.32 & 0.084 (0.005) & 0.087 & 0.082 & 0.072 \\
		38 &  & -252.97 & -209.39 & -302.1 &  & 0.083 & 0.079 & 0.069 \\
		39 &  & -207.76 & -162.43 & -264.85 &  & 0.075 & 0.077 & 0.066 \\
		40 & -118.72 (0.2) & -170.0 & -119.41 & -230.47 & 0.077 (0.01) & 0.071 & 0.071 & 0.063 \\
		41 & -83.52 (0.5) & -131.05 & -82.96 & -198.85 & 0.074 (0.01) & 0.074 & 0.066 & 0.059 \\
		42 & -51.48 (0.5) & -90.31 & -49.96 & -169.97 & 0.07 (0.01) & 0.073 & 0.062 & 0.056 \\
		43 & -23.72 (0.5) & -49.89 & -21.99 & -143.79 & 0.067 (0.02) & 0.072 & 0.054 & 0.053 \\
		44 &  & -11.78 & -2.43 & -120.25 &  & 0.067 & 0.045 & 0.049 \\
		45 &  & 16.95 & 13.89 & -99.26 &  & 0.052 & 0.046 & 0.046 \\
		\hline \hline
	\end{tabular}
	\caption{Full comparison of 3 potentials for the $C \ ^1\Sigma^+$ state in LiRb, all numbers are in cm$^{-1}$.  Blank entries denote vibrational levels for which there is no exp data or there is a flaw in the exp data.  The number in parenthesis is the error in the experimental value.}
	\label{tab:Ccomparison}
\end{table*}

\end{document}